\documentstyle[aps,epsf]{revtex}  

\begin{document}        

\baselineskip 14pt
\title{Spin Structure Functions $g_1$ and $g_2$ for the Proton and Deuteron}
\author{Gregory S. Mitchell (Representing the E155 Collaboration)}
\address{University of Wisconsin-Madison, 1150 University Avenue, Madison, WI 53706-1390}
%
\maketitle              

\begin{abstract}        
The experiment E155 at SLAC measured the spin structure functions 
$g_1$ and $g_2$ of the proton and deuteron.  The experiment used deep 
inelastic scattering of 48.3 GeV longitudinally polarized electrons 
incident on polarized solid $^{15}$NH$_3$ and $^6$LiD targets.  The data 
taken by three independent spectrometers covered a kinematic range of 
0.014 $< x <$ 0.9 and 1 (GeV/c)$^2 < Q^2 < $ 40 (GeV/c)$^2$.  Due to the 
high luminosity and polarization available at SLAC the data on $g_1$ are 
to date the most precise in this kinematic range.  The $x$ and $Q^2$ 
dependence of $g_1$ has been studied using NLO PQCD fits, allowing extraction
of values for the Bjorken sum rule and quark and gluon spin 
contributions to the nucleon.  Results are presented for $g_1$
and $g_2$ for the proton and deuteron.
\
\end{abstract}   	

\section{Introduction}               

Polarized deep inelastic scattering is a powerful tool for 
studying the internal spin structure of the nucleon~\cite{Roberts}. 
Early results obtained by experiments at SLAC~\cite{E80,E130} and CERN~\cite{EMC}
indicated that quarks contribute very little to the nucleon's spin.  This
contradicted expectations of the naive quark-parton model,
and led to the so-called spin crisis.  Additional theoretical 
and experimental studies followed,
resulting in a richer view of the nucleon. Experiments at 
CERN~\cite{SMC,SMC-nlo}, 
SLAC~\cite{E142,E143,E154,E154-nlo,E155}, and DESY~\cite{HERMES} have
provided increasing levels of precision and enabled the determination
of polarized parton distribution functions using next-to-leading order
perturbative QCD calculations and the (Altarelli-Parisi or DGLAP) evolution 
equations~\cite{SMC-nlo,E154-nlo}.  The main goal of the E155 experiment
at SLAC was, for both the proton and  deuteron, 
to make a precision measurement of $g_1$ covering a wide kinematic range.

\section{Inclusive Deep Inelastic Scattering and Polarized Structure Functions}

To study the internal spin structure of the nucleons,
the SLAC experiments have used the inclusive deep inelastic scattering
of a polarized electron beam incident on a fixed polarized nucleon target.
The kinematics of a scattering event are determined by the incident
electron energy ($E_0$) and the energy ($E^{\prime}$) and 
angle ($\theta$) of the scattered electron with respect to the incident
beam direction.
The event kinematics are described by the 4-momentum transfer squared 
$Q^{2}=4 E_0 E^\prime \sin^2\left( \theta / 2 \right)$, 
and the Bjorken scaling variable 
$x={Q^2 \over 2 M \nu}= {Q^2 \over  2 M (E_0-E^\prime)}$.  
The Bjorken scaling variable gives the fraction of the nucleon
momentum carried by the struck quark.  

In the unpolarized case, the cross-section for 
inclusive measurements is related to the unpolarized
structure functions $F_1$ and $F_2$.
In the polarized case, the difference between cross-sections for
anti-aligned versus aligned electron spins (denoted $\uparrow$) 
and nucleon spins (denoted $\Uparrow$) is given by
\begin{equation}
{d^2 \sigma \over d\Omega dE'}^{\downarrow\Uparrow}
-
{d^2 \sigma \over d\Omega dE'}^{\uparrow\Uparrow}
=
{4 \alpha^2 {E'} \over Q^2 E_0 M \nu}
\left[
(E_0 + E'\cos\theta) g_1(x,Q^2) 
-
2 x M g_2(x,Q^2)
\right]
\label{eq:parallel_diff}
\end{equation}
for the case of longitudinal polarization of the target, and by
\begin{equation}
{d^2 \sigma \over d\Omega dE'}^{\downarrow\Leftarrow}
-
{d^2 \sigma \over d\Omega dE'}^{\uparrow\Leftarrow}
=
{4 \alpha^2 {E'} \over Q^2 E_0 M \nu} \sin \theta
\left[
g_1(x,Q^2) 
+
{2 E_0 \over \nu} g_2(x,Q^2)
\right]
\label{eq:perp_diff}
\end{equation}
for the case of transverse polarization of the target,
where $M$ is the nucleon mass and $\alpha$ is the fine structure constant.
The notation
$\sigma^{\downarrow \Uparrow} = {d^2 \sigma \over d\Omega dE'}^{\downarrow\Uparrow}$
(and similarly for other cases) is used below.

Equations~(\ref{eq:parallel_diff}) and~(\ref{eq:perp_diff}) introduce
the spin structure functions $g_1(x,Q^2)$ and $g_2(x,Q^2)$.
These functions differ for the various targets 
(proton $p$, neutron $n$, and deuteron $d$), but are related by
$g_1^d = {1 \over 2}(g_1^p + g_1^n) (1-1.5 \omega_D)\,$.
The factor $(1-1.5\omega_D)$ corrects for the D-state
probability of the deuteron, $\omega_D$~=~0.05$\pm$0.01.
This probability is treated as a constant with respect to Bjorken $x$.

Since in both cases on the left hand side 
of Eqs.~(\ref{eq:parallel_diff}) and~(\ref{eq:perp_diff})
the cross-sections are nearly equal, to measure $g_1$
and $g_2$ by measuring the cross-sections and taking the differences
would require detailed and accurate knowledge of detector
acceptance and efficiency.  Instead, the spin structure functions
are obtained by measuring asymmetries, where these common factors divide out.

Measuring cross-section asymmetries $A_\|$ and $A_\bot$,
\begin{equation}
A_\|  =  \frac{\sigma^{\downarrow \Uparrow} 
                    - \sigma^{\uparrow \Uparrow}}{
                    \sigma^{\downarrow \Uparrow} 
                    + \sigma^{\uparrow \Uparrow}} 
\hspace{3em}{\mathrm and}\hspace{3em} 
A_\bot  =  \frac{\sigma^{\downarrow \Leftarrow} 
                    - \sigma^{\uparrow \Leftarrow}}{
                    \sigma^{\downarrow \Leftarrow} 
                    + \sigma^{\uparrow \Leftarrow}}
\, ,
\end{equation}
yields the polarized structure functions via
\begin{equation}
\label{equation:g1andg2}
g_1 = {F_1 \over D^\prime} \left[ A_\| + A_\bot \tan{(\theta/2)} \right]
\hspace{3em}{\mathrm and}\hspace{3em} 
g_2 = {F_1 \over D^\prime} {y \over 2 \sin\theta} 
\left(- A_\| \sin\theta + A_\bot {E_0 + E^\prime \cos \theta \over E^\prime}\right) \, .
\end{equation}
The quantities $g_1$, $g_2$, $F_1$, $A_\|$, and $A_\bot$ above 
are all functions of $x$ and $Q^2$.  The kinematic variables used in the above 
expression are 
determined from $E_0$ and either $(x,Q^2)$ or $(E^\prime,\theta)$ as follows:
$y=\nu / E_0$, $z=xM/E_0$,
$\epsilon = 1/\left[ 1 + 2(1+\nu^2/Q^2) \tan^2(\theta/2)\right]$,
$\gamma^2 = 4M^2x^2/Q^2$, and
$D^\prime =(1-\epsilon)(2-y) / \left[ 1+\epsilon R(x,Q^2) \right]$.
$F_1$ was calculated from 
fits to world data on $F_2$ \cite{F2NMC} and $R$ \cite{E143-R1998}.

In leading order $g_1$ is only sensitive to the net quark polarizations. 
However, in next-to-leading order (NLO) perturbative QCD (PQCD)
$g_1$ is also sensitive to the polarized gluon distribution $\Delta G$ through
the $Q^{2}$ evolution of the polarized parton distributions.
This evolution is governed by the
Altarelli-Parisi (or DGLAP) evolution equations\cite{DGLAP} and
reflects the increasing resolution of deep inelastic scattering
with increasing $Q^{2}$.  One important consequence
of this evolution is to link the quark
distributions to the gluon distributions through g $\rightarrow$ gq
splitting.  This makes it possible to indirectly measure the 
gluon distributions by measuring the quark distributions over a
wide range in $Q^{2}$.  Given sufficient statistical 
power over a wide range in $x$ and $Q^{2}$, NLO PQCD 
analysis of the $Q^{2}$ dependence of $g_{1}$ can be used to constrain the 
net polarization of the gluon.  

Although the naive picture of the $g_2$ structure function is unclear,
twist-2 calculations from Wandzura and Wilczek~\cite{g2ww} 
provide a relation between $g_2$ and $g_1$: 
\begin{equation}
g_{2}^{WW}\!(x,Q^{2}) = -g_{1}(x,Q^{2}) + \int^1_x g_{1}(\zeta,Q^{2})d\zeta/\zeta \, .
\label{EQ:g2ww}
\end{equation}
More generally, $g_2(x,Q^2)$ can be written as
$g_2(x,Q^2) = g_2^{WW}\!(x,Q^2) - \int_x^1{\partial \over \partial y}
\left({m\over M}h_T(y,Q^2)+\xi(y,Q^2)\right){dy\over y} \,$,
where the second and third terms are twist-2 and twist-3, respectively.
\par
\section{The Experiment}
The experiment E155 ran in early 1997 and collected $\sim$180 million
deep inelastic scattering electron events.  Keys to the success of E155 included:
the high intensity, highly-polarized  beam; the use of lithium deuteride as
the deuteron target material; and the construction of a third spectrometer.

Longitudinally polarized 48.3 GeV electron beam pulses \cite{PolSource}
of up to 400 ns duration were produced at 120~Hz by a 
circularly polarized laser beam illuminating a strained GaAs photocathode.  
The beam polarization $P_{b}$=0.810~$\pm$~0.020  was determined 
using M{\o}ller scattering from 20--154 $\mu$m thick Fe-Co-V 
polarized foils~\cite{Moller}.

Two polarized nucleon target materials were used in E155: 
ammonia ($^{15}$NH$_{3}$) as a proton target; and lithium deuteride 
($^6$Li$^2$H, or $^6$LiD)~\cite{LiD} as a deuteron target.
As in E143~\cite{Target},
polarization of the target materials was obtained using the technique of 
dynamic nuclear polarization (DNP). 
In this technique, a combination of microwaves ($\sim$140~GHz), low temperature 
(1~K), and a high magnetic field (5~T) polarizes paramagnetic
electron centers in the target material, and transfers that polarization to 
the nucleons.
Nuclear magnetic resonance (NMR) measurements were made using coils
embedded in the target materials to determine the polarization at
regular intervals.  These measurements were calibrated to the
signal measured at thermal equilibrium near 1.6 K~\cite{MeyerCrabb}.
The polarizations slowly decreased over time due to
radiation damage, and were restored by periodic annealing at about
80~K for the NH$_3$ target and 185~K for the LiD.  
As compared to deuterated ammonia ($^{15}$ND$_{3}$), 
lithium deuteride provides a larger ratio of polarized (effective) deuterons to
the total number of nucleons as well as higher radiation resistance.  The 
$^6$Li can, to first order, be treated as a polarized deuteron plus an 
unpolarized alpha particle, and therefore half of the nucleons in $^6$LiD
are the desired polarizable species.  Average polarizations of
$\langle P_{t} \rangle \approx$ 80\% for the proton target and
$\langle P_{t} \rangle \approx$ 22\% for the deuteron target were achieved.
Overall relative uncertainties on P$_{t}$ of 6\% (preliminary) and 4\% were 
obtained for the proton and deuteron respectively.

\begin{figure}[htb]	
\centerline{
\hbox{\epsfxsize=7.truein \epsfbox{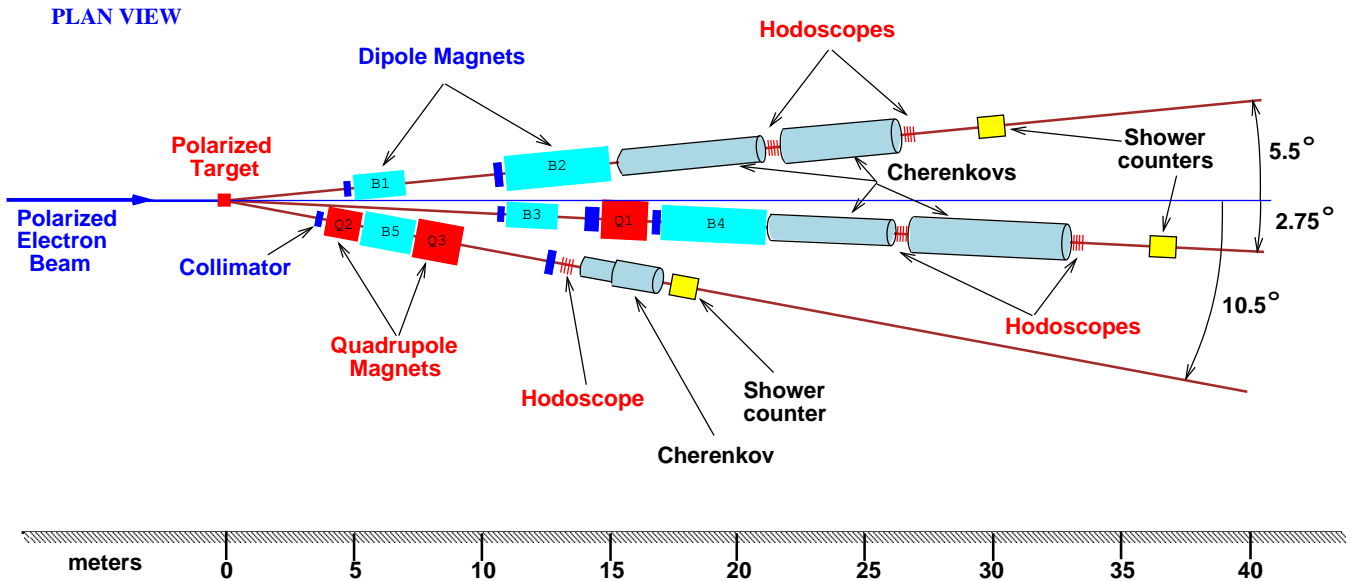}}
}
\vskip .1cm
\caption[]{
\label{fig:spectrometers}
\small The E155 spectrometers. }
\end{figure}

Scattered electrons were detected in three independent magnetic spectrometers 
at central angles of 2.75$^\circ$, 5.5$^\circ$, and 10.5$^\circ$ with
respect to the incident beam, as shown in Fig.~\ref{fig:spectrometers}.
The spectrometers at 2.75$^{\circ}$ and 5.5$^{\circ}$ were used
previously in the experiment E154~\cite{E154}.  
In each spectrometer, electrons were identified by threshold gas Cherenkov 
counters and a total absorbing lead glass electromagnetic calorimeter.
Particle momenta and scattering angles were measured 
with sets of scintillator hodoscopes.  The 10.5$^\circ$  
spectrometer was added for E155 to double the $Q^{2}$ range of
the experiment.

\section{Experimental Asymmetries}
The experimental asymmetries were determined
from the numbers of scattered electrons per incident beam charge
for negative and positive beam helicity for each target, divided
by the beam and target polarizations.
The rates were corrected for
contributions from charge symmetric background processes, which 
were measured by reversing the 
spectrometer polarities. Corrections were also included for
mis-identified hadrons, which were typically 2\% or 
less of the electron candidates.  
Also, a dilution factor accounted for the fraction of events 
originating from the polarizable protons or free deuterons,
as opposed to the other materials in the target. 
The average dilution factor for the proton target was 0.15, for
the deuteron target 0.19.

Nuclear correction factors accounted for the
presence of several polarizable nuclear species in the target. 
For the proton, a small correction was made for 
the polarization of the nitrogen nuclei.  
For the deuteron, the measured asymmetry included 
contributions from the free deuterons and effective deuterons in $^{6}$Li.
The effective deuteron in $^6$Li
has a net polarization of 87\% of the $^6$Li polarization~\cite{Rondon},
which leads to an effective dilution factor of 
$\sim$0.36 for $^{6}$LiD, as compared with $\sim$0.22 for $^{15}$ND$_{3}$. 
The measured deuteron asymmetries also were corrected for small 
contributions from polarized protons in Li$^{1}$H and $^7$Li.

Both internal \cite{RCinternal} and external \cite{RCexternal}
radiative corrections were obtained using an iterative global fit of
all data, including E155. Previous SLAC data were recorrected in a 
manner consistent with the E155 corrections. 

\section{Results}
\subsection{Longitudinal Structure Function $g_1$}

\begin{figure}[htb]	
\centerline{
\hbox{\epsfxsize 4.0 truein \epsfbox{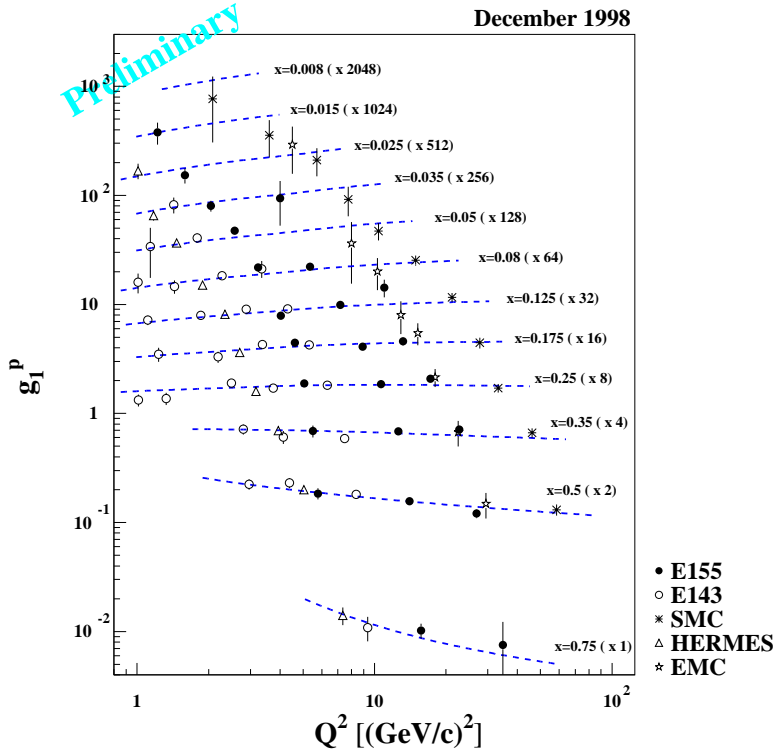}}
}
\vskip .3cm
\caption[]{
\label{fig:fanplot}
\small 
$g_1^p$ for E155 and other experiments vs. $Q^2$.
The dashed curves are from a functional fit to world $g_1$ data.
The points and curves have been scaled by the factors
in parentheses for clarity.
}
\end{figure}

The E155 proton results are plotted as 
$g_1^p$ vs.$\;$$Q^2$ in Fig.~\ref{fig:fanplot}.
There is a remarkable agreement between the many experiments.
The world data on $g_1^p$ exhibit a clear $Q^2$ dependence, and the
scaling violation is similar in character to that of the unpolarized structure
functions.
At low $x$, the spin structure function increases with $Q^2$, and at high
$x$ it decreases with increasing $Q^2$.

\begin{figure}[htb]	
\centerline{
\hbox{\epsfxsize 4.0 truein \epsfbox{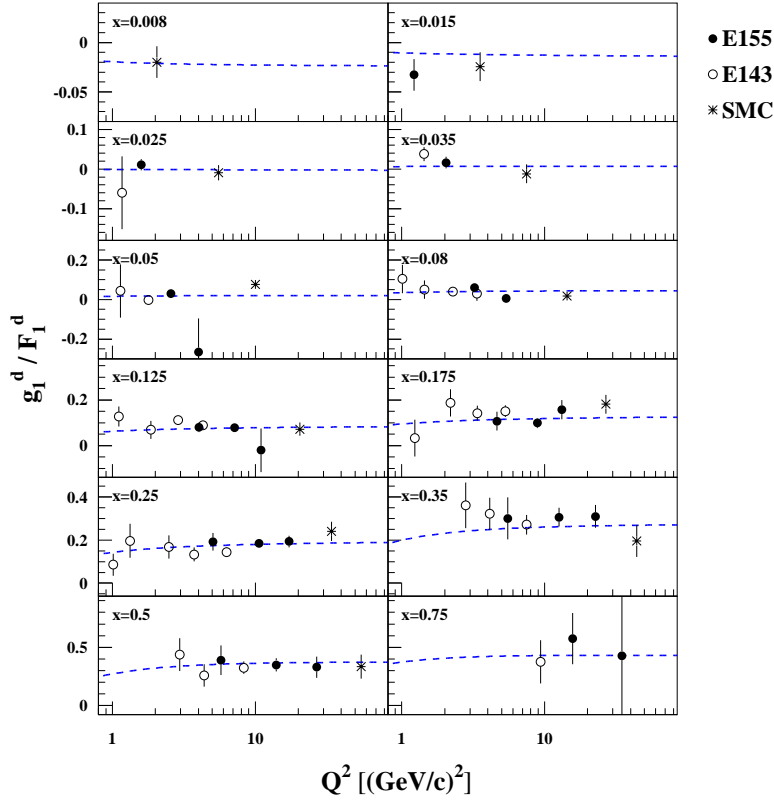}}
}
\vskip .3cm
\caption[]{
\label{fig:q2dep}
\small 
${g_{1}}^d/{F_{1}}^d$ for E155 vs. other experiments. 
The multiple points in each bin for E155 are for the separate spectrometers.
The dashed curves are from a functional fit to world data on $g_1/F_1$.
}
\end{figure}

The E155 deuteron results~\cite{E155} are presented as $g_{1}^{d}$/$F_{1}^{d}$ 
vs.$\;$$Q^{2}$ in Fig.~\ref{fig:q2dep}.
They are also in good agreement with world data, and the data from the
three spectrometers provide both large $Q^{2}$ coverage and
good statistical resolution in the mid-$x$ region.
There is no significant $Q^{2}$ dependence
for $g_{1}^{d}/F_{1}^{d}$, and likewise $g_{1}^{p}/F_{1}^{p}$,
indicating that the polarized and 
unpolarized structure functions evolve similarly.  

Evolving the data to a common $Q^2$ by use of a functional fit to 
world $g_1$ data, and integrating over the kinematic $x$ range of E155 
yields preliminary results for $Q^2=5$~(GeV/c)$^2$ : 
$\int_{0.014}^{0.9} {g_1}^p \ dx  =  0.132  \pm 0.002 \pm 0.010\;$
 and 
$\int_{0.014}^{0.9} {g_1}^d \ dx  =  0.044  \pm 0.003 \pm 0.003\,$,
where the first error is statistical and the second is systematic. 
The extrapolations for the unmeasured high $x$ region are negligible.
However, the contribution from the low $x$ region below 0.014 
does not converge for the functional fit, reinforcing the need for 
additional data at very low $x$. 
Using the E154 \cite{E154-nlo} PQCD fit
to calculate the contribution for the low $x$ region gives values: 
$\int_0^1 g_1^p(x,Q^{2})\;dx=\Gamma_1^p=0.126\pm0.003\pm0.010\pm0.009$
and 
$\int_0^1 g_1^d(x,Q^{2})\;dx=\Gamma_1^d=0.030\pm0.005\pm0.004\pm0.005$,
where again the first error is statistical, the second is
systematic, and the third error is a theory error.
These values can be combined to obtain a preliminary E155 
result for the Bjorken sum rule~\cite{Bjorken},
$\Gamma_1^{p-n} = \int_{0}^{1} {g_1}^p - {g_1}^n = 
\int_{0}^{1} {g_1}^p - ( {g_1}^d {2 \over (1-1.5 \omega_D)} - {g_1}^p)
= 0.187 \pm 0.012 \pm 0.022 \pm 0.021 $.
This is consistent with the theoretical prediction at $Q^2=5$~(GeV/c)$^2$, 
$\Gamma_1^{p-n} = 0.182 \pm 0.005$,
which includes third order QCD corrections~\cite{Larin}, 
using $\alpha_s(M_Z^2)=0.119\pm0.002$~\cite{PDG}.

Using the published NLO PQCD fit methods of E154~\cite{E154-nlo}, but 
including final SMC results~\cite{SMC} and preliminary E155 results in
the data set, polarized parton distributions were extracted.
The eight-parameter fit had a $\chi^2$/d.o.f. of 206/210.
From the parton distributions, the following preliminary 
integral results are obtained at $Q^2=5$~(GeV/c)$^2$, 
in the $\overline{\mathrm{MS}}$ scheme:
$\Gamma_1^{p}$ =  $0.116 \pm 0.005 \pm 0.009$,
$\Gamma_1^{d}$ =  $0.028 \pm 0.004 \pm 0.007$, and
$\Gamma_1^{p-n}$ =  $0.172 \pm 0.005 \pm 0.008$,
where the first errors are statistical and the second systematic,
and similarly sized theory errors are ignored. 
The Bjorken sum rule result is again consistent with the theoretical prediction.
Additionally, the fit polarized parton distributions yield the preliminary 
results:
${\Delta}G$ = $1.8 \pm 0.6 \pm 1.3$
and $\Delta\Sigma$ = $0.22 \pm 0.04 \pm 0.06$.
The data indicate that the gluon
contribution to nucleon spin $\Delta G$ is positive, but do not constrain
the value well.  The data do well constrain the quark spin contribution 
$\Delta \Sigma$, and it is smaller than  predicted naively or by 
the Ellis-Jaffe sum rule~\cite{EllisJaffe}.

\subsection{Transverse Structure Function $g_2$}
\begin{figure}[htb]	
\centerline{
\hbox{\epsfxsize 6.2 truein \epsfbox{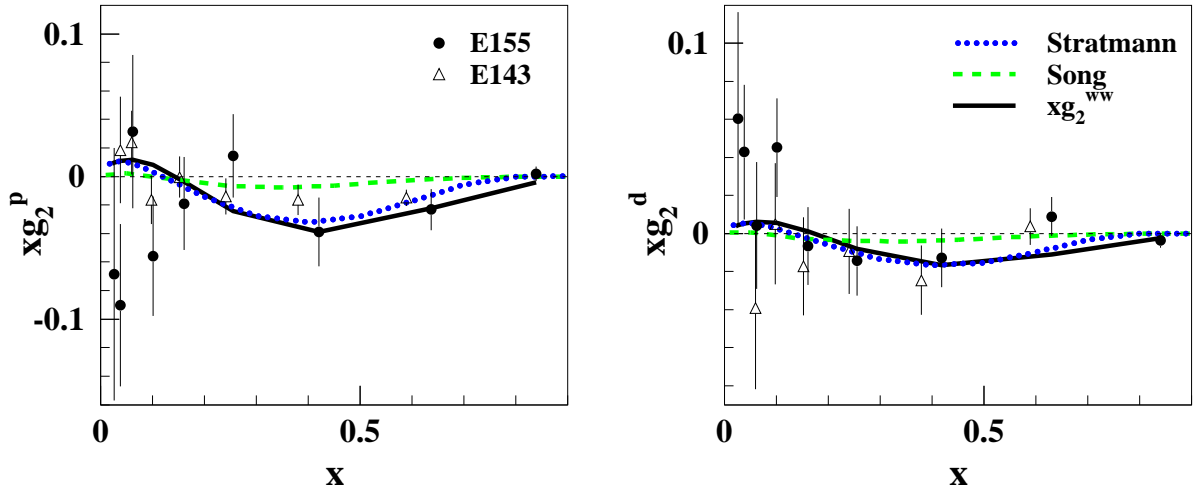}}
}
\vskip .3cm
\caption[]{
\label{fig:xg2}
\small 
Plot of $xg_2$ for proton and deuteron.
The error bars are statistical.  The solid line indicates 
the prediction from Wandzura and Wilczek 
\protect\cite{g2ww} using a fit to world $g_1$ data.  Also 
shown are calculations from Stratmann~\cite{Stratmann} and Song~\cite{Song}.
}
\end{figure}

E155 ran for a short period during the 1997 run at 38.8 GeV beam
energy with the target polarization transverse to the beam
direction.  Results for the $g_2$ structure function obtained from this
data~\cite{E155} are shown (as $xg_2$) in Fig~\ref{fig:xg2}.  
The data from the three spectrometers show no $Q^{2}$ dependence 
to the virtual photon-nucleon asymmetry $A_2$ within uncertainties, 
so the data from the individual spectrometers have been combined.  
The data are consistent with $g_2^{WW}\!$, and satisfy the 
Burkhardt-Cottingham sum rule~\cite{Burkhardt} within errors.  However,
the current results lack the power to differentiate
among $g_2^{WW}\!$, model predictions
\cite{Stratmann,Song}, and $g_2(x,Q^2)=0$.
This will be addressed by the much larger data set planned
for the E155 extension run (March-May 1999), 
which will focus on measurements
with the target polarization perpendicular to the beam direction
in order to make a high precision measurement of the $g_2$ structure function.
The data from this run will significantly improve 
knowledge of $g_2$ and will definitively distinguish between
$g_2$~=~0 and $g_2^{WW}\!$. 

\end{document}